\documentstyle[12pt,twoside,fleqn,espcrc2]{article}

\newcommand{\AmS}{{\protect\the\textfont2
A\kern-.1667em\lower.5ex\hbox{M}\kern-.125emS}}
\title{Confinement and monopole condensation: some properties of the
disorder parameter.}
\author{A. Di Giacomo$^{\rm a}$\thanks{
Speaker at the Conference.
Partially supported by MURST and by EC, FMRX-CT97-0122}
and G. Paffuti\address{Dip. Fisica
Universit\`a and INFN Pisa\\
Via Buonarroti 2, 56125 Pisa}%
}
\begin{document}
\maketitle
\begin{abstract}
We demonstrate that the disorder parameter $\langle\mu\rangle$ used in
\cite{1} to detect dual superconductivity in the confining phase of QCD
is
the v.e.v.of a magnetically charged, Dirac like, gauge invariant
operator
$\mu$. We also show that the abelian projection on the lattice is
determined up to terms ${\cal O}(a^2)$ ($a =$ lattice spacing).
\end{abstract}
\section{Introduction}
We consider a $U(1)$ gauge theory with a conserved magnetic current
$j^M_\mu
= \partial^\nu F^*_{\mu\nu}$. The corresponding magnetic $U(1)$ symmetry
can either be Wigner or broken \`a la Higgs. In the first case (Wigner)
the Hilbert
space consists of superselected sectors with definite magnetic charge.
In the second
case (Higgs) at least one magnetically charged operator $\mu$ exists,
with $\langle
\mu\rangle\neq 0$. $\langle\mu\rangle$ is the order parameter. The free
energy density
(effective Lagrangean), is uniquely determined by symmetry and
dimensional
arguments\cite{2} to be
\begin{equation}
{\cal L} = \left(D_\rho\langle\mu\rangle\right)^*
D_\rho\langle\mu\rangle - \frac{1}{4} F^{\mu\nu} F_{\mu\nu} -
V(\langle\mu\rangle)
\end{equation}
Here $D_\rho = \partial_\rho - i q_M \tilde A_\rho$ is the dual
covariant derivative,
$\tilde A_\rho$ the dual vector potential, $V(\langle\mu\rangle)$ the
usual quartic
potential. If $V$ has a non trivial minimum, $\langle\mu\rangle\neq 0$,
the system
is a dual superconductor. In the usual non compact formulation
$j^M_\mu\equiv 0$
(Bianchi identities). On a lattice the theory is compact and a non zero
magnetic
current can be defined\cite{3}.
In QCD $U(1)$ is defined by abelian projection\cite{4}. A magnetically
charged operator
$\mu$ has been constructed\cite{1} and $\langle\mu\rangle$ has been used
as  a
disorder parameter for detecting dual superconductivity.
The results of this investigation are the following\cite{5}:
1)The confined phase of quenched QCD has $\langle\mu\rangle\neq
0$, and behaves as a dual superconductor in all the abelian
projections\cite{1}.
2)In the deconfined phase $\langle\mu\rangle = 0$ and superselected
magnetic
sectors exist.
3)In the vicinity of the transition
$
\langle\mu\rangle \mathop\propto_{T\to T_c^-} \left(1 -
\frac{T}{T_c}\right)^\delta
$.
$\delta$ is independet of the abelian projection, and is equal to the
analogous index
of the dual Polyakov line\cite{6}.
4)A similar behaviour is found in the presence of dynamical quarks.
These results are obtained by properly performing the infinite volume
limit, by use of
finite size scaling techniques\cite{1,5,6}, and provide basic
information on the dual
structure of QCD. The questions we want to address here are the
following\\
1)Does $\langle\mu\rangle \neq 0$ contradict the so called Elitzur's
theorem\cite{8}?
2)How precisely is the abelian projection definable on the lattice?
\section{The order parameter of a superconductor.}
The ground state of a superconductor is a superposition of states with
different
charges\cite{9}, and the order parameter $\langle\varphi\rangle$ is the
v.e.v. of a
charged operator $\varphi$.
{\bf{Theorem}\,} (continuum version of ref.\cite{8})
If $\varphi(x)$ is a local charged operator, in a gauge invariant
formulation (no
gauge fixing), $|0\rangle$ is gauge invariant and hence
\[
\langle0|\varphi(x)|0\rangle=
e^{i\Lambda(x)}\langle0|\varphi(x)|0\rangle\quad \forall \Lambda(x)
\]
or
$
\langle0|\varphi(x)|0\rangle = 0$.
Does the existence of a superconductor violate gauge invariance? The
answer is of
course no.
In the usual perturbative treatment of the Higgs phenomenon a gauge is
fixed, e.g. the
unitary gauge\cite{10}, and the theorem is eluded. In a gauge invariant
formulation,
like lattice, the way out is to define gauge invariant charged
operators\cite{11},
$\tilde\varphi(x)$, as follows
\begin{equation}
\tilde \varphi(x) = \varphi(x) e^{i (A,h)}\label{eq3}
\end{equation}
where
\[
(A,h) = \int d^4y A_\mu(y) h_\mu(y-x)\]
$\partial^\mu h_\mu(z) = \delta^4(z)$.
Under a generic gauge transformation $U_\Lambda$, with
$\Lambda(x)\to 0$ as ${|x|\to\infty}$,
$(A,h)\to (A,h) - \Lambda(x)$, $\varphi(x)\to e^{i\Lambda(x)}\varphi(x)$
and hence
$\tilde\varphi(x)\to\tilde\varphi(x)$. $\tilde\varphi(x)$ is gauge
invariant. However,
under a global transformation (with $\partial_\mu\Lambda = 0$) $A_\mu\to
A_\mu$
and $\tilde\varphi(x) \to e^{i\Lambda}\tilde\varphi(x)$ like any charged
operator.
$\tilde \varphi(x)$ is charged and gauge invariant. It is non local, but
by a
judicious choice of $h_\mu$, it obeys cluster property
\begin{equation}
G(x)\equiv \langle \tilde\varphi(x)^\dagger \tilde\varphi(0)\rangle
\mathop\simeq_{|x|\to \infty}
A e^{- M |x|}
+ |\langle\varphi\rangle|^2
\label{eq4}\end{equation}
and defines the order parameter $\langle\varphi\rangle$ by the
asymptotic behaviour.
Possible choices for $h_\mu$ depend on the dimension $d$ of its support.
For $d=1$
$h^\mu = \delta^\mu_0 \theta(x_0-y_0)\delta^3(\vec x - \vec y)$ and
$\exp(i(A,h) = \exp(i\int_{-\infty}^{x_0} A_0(y_0,\vec x) dy_0)$ is  a
parallel
transport from $-\infty$ along time axis (Mandelstam string). The string
can be put on
any path $C$ going to infinity and $\tilde \varphi(x) =
\varphi(x)\exp(i\int_C A_\mu
dx^\mu)$.
For $d=3$ $h^\mu = (0,\frac{1}{4\pi}\frac{\vec x - \vec y}{|\vec x -
\vec y|^3})$
(Dirac choice).
For a lattice version of the Higgs model
\begin{equation}
{\cal L} =\frac{\beta}{2} - a^2\cos(d\theta - e A)\label{eq5}
\end{equation}
it can be proved that\\
1)
At sufficiently small $a$ (Coulomb phase)
$|G(x)|\mathop\leq_{|x|\to\infty}C\exp(-\rho |x|) $
\\
2)
At sufficiently large $a$ and small $e$ (Higgs phase) $|G(x)|\geq a^2$
as
$|x|\to \infty$, provided $(h,h)$ is uniformely bounded in the limit
$V\to\infty$.
This is true for the Dirac choice of $h^\mu$ ($d=3$) but not for the
string ($d=1$).\\
3)The Hilbert space enlarged by the correlators of $\tilde\varphi$ obeys
Osterwalder-Schrader positivity, and admits a decomposition in
superselected sectors
in the Coulomb phase.
\section{Dual superconductivity.}
For compact $U(1)$ the operator $\mu$ which creates a monopole is
defined as\cite{12}
\[
\mu(x_0,\vec x) =
e^{
\beta\sum_{\vec n}
\left(\cos[\theta_{0i}(x_0,\vec n)-b_i(\vec x-\vec n)]
-\cos\theta_{0i}(x_0,\vec n)\right)
}\]
where $b_i(\vec x -\vec y)$ is the vector potential at $\vec y$ of a
monopole sitting
at $\vec x$.
Theorems\\
1)$\mu$ carries magnetic charge\cite{12}.
2)$\mu$ is gauge invariant ($\theta_{0i} = F_{0i}$).
3)$\mu$ is Dirac like\cite{13,14}.
4)The correlations functions  $\langle \mu(x)\bar\mu(0)\rangle$ are
equal to
$\langle\tilde\varphi(x)\tilde\varphi^\dagger(0)\rangle$ of the Higgs
model with
the change $\beta_H \to \beta = \frac{1}{4\pi^2\beta_H}$, as required by
duality.
In conclusion $\mu$ is a gauge invariant charged operator and
$\langle\mu\rangle$ a
legitimate disorder parameter for dual superconductivity.
\section{About the abelian projection.}
The abelian projection is a gauge transformation which diagonalizes an
operator
$\Phi(x) = \sum\Phi^a T^a$ in the adjoint representation. In this gauge
the generic
link on the lattice $U_\mu(n)$ can be written in the form
\begin{equation}
U_\mu(n) = V_\nu(n) C_\mu(n)\label{eq7}\end{equation}
where $C_\mu(n)$ is an exponent of the diagonal generators (photons),
$V_\mu(n)$ of
the off diagonal (charged) generators. $C_\mu$ is uniquely defined. Also
the form
\[ U_\mu(n) = C_\mu(n) {V'}_\mu(n) \]
is possible, with $ V'_\mu = C^\dagger(n) V_\mu C_\mu(n)$, again a
charged operator.
For the plaquette $\Pi_{\mu\nu} = U_\mu(n) U_\nu(n+\hat\mu)
U_\mu^\dagger(n+\hat\nu)
U^\dagger_\nu(n)$ the abelian projection is not uniquely defined. It can
be defined,
e.g., as the projected part of $\Pi_{\mu\nu}$ by a procedure similar to
that leading
to eq.(1), or, as usually done, as the plaquette constructed with the
abelian links
$\Pi^0_{\mu\nu} = C_\mu(n) C_\nu(n+\hat\mu) C_\mu^\dagger(n+\hat\nu)
C^\dagger_\nu(n)$. By use of eq.(7)
\begin{eqnarray}
&&\hskip-10pt
\Pi_{\mu\nu} = \label{eq8}\\
&&\hskip-10pt V_\mu(n) V'_\nu(n+\hat\mu) (V_\mu^{''})^\dagger(n+\hat\nu)
(V_\nu^{'''})^\dagger(n) \Pi^0_{\mu\nu}
\nonumber
\end{eqnarray}
The $V$'s in eq.(8) are separately charged, but their product contains
at the exponent
terms coming from the commutators in the Baker-haussdorf formula, which
are ${\cal
O}(a^2)$ and belong to the diagonal subspace of the algebra.
The abelian projection on the lattice is undefined by terms ${\cal
O}(a^2)$.

\end{document}